\documentclass{elsart}

\usepackage{amssymb}

\newcommand{\adt}{\dot{a}}
\newcommand{\bdt}{\dot{b}}
\newcommand{\cdt}{\dot{c}}
\newcommand{\ddt}{\dot{d}}
\newcommand{\edt}{\dot{e}}
\newcommand{\fdt}{\dot{f}}
\newcommand{\gdt}{\dot{g}}

\newcommand{\pa}{\partial}

\newcommand{\abar}{\bar{a}}
\newcommand{\bbar}{\bar{b}}
\newcommand{\cbar}{\bar{c}}
\newcommand{\dbar}{\bar{d}}
\newcommand{\ebar}{\bar{e}}

\newcommand{\gbar}{\bar{g}}

\newcommand{\Tr}{\,\mathrm{Tr}} 

\newcommand{\calR}{\mathcal{R}}
\newcommand{\lag}{\mathcal{L}}
\newcommand{\ad}{\mathrm{ad}}
\newcommand{\GamHat}{\hat{\Gamma}}
\newcommand{\iso}{\simeq}

\begin{document}

\begin{flushright}
\hfill{AEI-2003-089}\\
\hfill{hep-th/0310256}
\end{flushright}       

\begin{frontmatter}

\title{Hidden symmetries in minimal five-dimensional supergravity}

\author[al1]{Markus P\"ossel} and 
\ead{mpoessel@aei.mpg.de}
\author[al1]{Sebastian Silva}
\address[al1]{Albert-Einstein-Institute\\
              Max-Planck-Institute for Gravitational Physics\\
              Am M\"uhlenberg 1,
              D-14476 Golm,
              Germany}

\begin{abstract}
We study the hidden symmetries 
arising in the dimensional reduction of $d=5$, $\mathcal{N}=2$ supergravity to 
three dimensions.
Extending previous partial results for the bosonic
part, we give a derivation that includes fermionic terms, shedding  
light on the appearance of the local 
hidden symmetry $SO(4)$ in the reduction.
\end{abstract}

\end{frontmatter}

\section{Introduction}
\label{Intro}

Since their role in the greater scheme of M-theory was postulated
\cite{Hull:1995ys}, there has been intense
renewed interest in the hidden symmetries of supergravity
\cite{Cremmer:1979up,Cremmer:1978ds,Cremmer:1980gs,Julia:1980gr}.
While most work has, understandably, centered on the maximal
eleven-dimensional theory and its various dimensional reductions, 
with recent work ranging
from the gauging of subgroups of the global exceptional groups 
\cite{Nicolai:2001sv,deWit:2002vt,Andrianopoli:2002mf} 
to the identification of new vacua
  \cite{Khavaev:1998fb,Fischbacher:2002fx} to possibilities for enlarging
the hidden symmetries even further  \cite{Koepsell:2002sg}, there have
also been results of a more general scope; notably the work showing 
in the most systematic manner yet how the hidden symmetries arise in
successive dimensional reduction \cite{Cremmer:1998ct,Cremmer:1998px}
has been generalized to other dimensional reductions to
three dimensions,  \cite{Cremmer:1999du}.

The present work, whose results form part of the thesis \cite{Possel:2003rk}, 
is concerned with the hidden symmetry arising from the reduction to
three dimensions of (minimal) five-dimensional 
supergravity \cite{Cremmer:1980gs,Chamseddine:1980sp}.  Since its inception,
that model, in many ways a ``little brother'' to the eleven-dimensional 
theory, has been used time and again to learn more about its 
higher-dimensional kin.  Recent examples include toy models of the M5 brane  
\cite{Boyarsky:2002ck}, cosmological models  \cite{Chen:1999eh},
and methods developed for the study of the U-dualities of 
M-theory  \cite{Mizoguchi:1999fu,Schroder:2000bu}.  
Concerning the hidden symmetry in question, there are so far only partial 
results, namely a construction \cite{Mizoguchi:1998wv} of the bosonic 
part of the model using a decomposition of $G_{2(+2)}$ with respect to 
$SL(3,\mathbb{R})$ and a corresponding construction by Cremmer et al.\ as part
of the aforementioned more general study of reductions to three dimensions
 \cite{Cremmer:1999du}.  What has been missing, so far, is a complete
analysis, notably one 
that includes the fermionic sector.  The latter is
interesting not only as another
data point in an area for which, in contrast to the bosons, no systematic 
scheme yet exists, namely the relationship between the local extended 
symmetry and the dimensional reduction of the spinors, but also 
for another reason:  In maximal supergravity,  
hidden symmetries have been successfully ``lifted'' to eleven
dimensions \cite{deWit:1986mz,Nicolai:1987jk,Drabant:1989bk}, and
recent work has uncovered tantalizing hints of ``exceptional geometric
structures'' associated with these liftings
\cite{Koepsell:2000xg,Koepsell:2001ta}.  The fermionic sector plays a 
crucial part in this type of lifting, and the results of this paper are thus
a prerequisite for a search for such ``exceptional geometry''
in five-dimensional
gravity.  The present paper contains a construction
of the $\mathfrak{g}_{2(+2)}/\mathfrak{so}(4)$ target model in three
dimensions and then proceeds to a dimensional reduction of five-dimensional
supergravity, including the fermionic sector, in which the emergence of
the hidden symmetry is shown.

\section{The {\boldmath\mbox{$\mathfrak{g}_{2(+2)}$}}-model 
in $\mathbf{(2+1)}$ dimensions}
\label{3DSection}

In this part, we construct the 
$\mathfrak{g}_{2(+2)}/\mathfrak{so}(4)$-supergravity
in $2+1$ spacetime dimensions.  For the sigma-model part we use the 
conventions\footnote{As for space-time conventions, $\mu,\nu,\dots$ are
curved and $\alpha,\beta,\dots$ flat space-time indices; our metric is 
``mostly plus''; our gamma matrices are real 
with $\gamma^0\gamma^1\gamma^2=+1$.}
(as well as some general formulae) of Marcus and 
Schwarz \cite{Marcus:1983hb}.  The basic fields of our model will be,
firstly, scalars $\varphi_i$ 
parametrizing the coset; as detailed in \cite{Marcus:1983hb},
they occur in the Lagrangian in the form of a Lie-algebra valued field
$P_{\mu}$ and composite connection coefficients $Q_{\mu}$;    
secondly, a dreibein field $e_{\mu}{}^{\alpha}$; finally,
fermionic superpartners of these bosonic fields: a spin-1/2 field
$\chi$ and a spin-3/2 (gravitino) field $\Psi$, respectively.  To match
degrees of freedom, we need an $\mathcal{N}=4$ extended supersymmetry
so that, in addition to Lorentz symmetry, the gravitino transforms 
non-trivially under an additional R-symmetry.

Before the actual construction, we need to assign the different
fields to their proper representations with respect to the internal 
symmetries involved, notably the local $\mathfrak{so}(4)$ symmetry of
our model.  In parallel with the $\mathfrak{e}_{8(+8)}$-case, one might
think that the $\mathfrak{so}(4)$ R-symmetry would have to be 
identified with the local $\mathfrak{so}(4)$ symmetry from the coset
construction; however, the situation is more complicated.  If we
assign the supersymmetry parameter $\epsilon$ (and hence the gravitino
$\Psi$) to the 
vector representation of the R-symmetry $\mathfrak{so}(4)$, then from
a general analysis of the supercharges as in \cite{deWit:1993up} we must 
conclude that regarding the representations possible for the 
(massless) matter fields, one chiral subgroup, henceforth denoted
$\mathfrak{so}(3)_F$, acts only on fermionic degrees on freedom, the
other, $\mathfrak{so}(3)_B$, only on bosons.  On the other hand, from
the group theory literature \cite{McKayTables} we know that the proper 
coset decomposition for the $\mathbf{14}$ representation of the
$\mathfrak{g}_2$ is $\mathbf{14}=(\mathbf{1},\mathbf{3})
+(\mathbf{3},\mathbf{1})+(\mathbf{4},\mathbf{2})$ (with the usual convention
of denoting representations by their dimensions, and with the tupels on 
the right hand side referring to the two chiral $\mathfrak{so}(3)$
components), so in particular, the coset scalars transform non-trivially
under both of the local $\mathfrak{so}(3)$.  To resolve the problem, we
need to introduce a third algebra, which we shall call $\mathfrak{so}(3)_2$.
This symmetry included, there is indeed an assignment to representations
of $\mathfrak{so}(3)_F\times\mathfrak{so}(3)_B\times\mathfrak{so}(3)_2$
that is consistent with the above requirements as well as with the 
form of the supersymmetry transformations (schematically,
$\delta_S\Psi\sim D\epsilon; \delta_S\chi\sim \varphi\epsilon;
\delta_S\varphi\sim \bar{\epsilon}\chi$), namely with $\Psi/\epsilon$
in the representation $(\mathbf{2_F},\mathbf{2_B},\mathbf{1_2})$,
the $\varphi$ transforming as $(\mathbf{1_F},\mathbf{2_B},\mathbf{4_2})$,
and the $\chi$ as $(\mathbf{2_F},\mathbf{1_B},\mathbf{4_2})$.

Next, we need the algebra for the coset decomposition.  We denote the indices
of the two chiral components $\mathfrak{so}(3)_B$ and $\mathfrak{so}(3)_2$
of the maximal compact $\mathfrak{g}_2$ subalgebra $\mathfrak{so}(4)$ by
$\bar{a},\bar{b},\ldots$ and $\dot{a},\dot{b},\ldots$, respectively.
Decomposed 
with respect to representations of that subalgebra, an algebra element
can be written as a contraction of coefficients with generators $E$, with the
part in $\mathfrak{so}(3)_2$ given as $M^{\adt}{}_{\bdt} E^{\bdt}{}_{\adt}$,
the $\mathfrak{so}(3)_B$ as $N^{\abar}{}_{\bbar} E^{\abar}{}_{\bbar}$
and the non-compact part as 
$Y^{{\abar}{\adt}{\bdt}{\cdt}}E_{{\abar}{\adt}{\bdt}{\cdt}}$.
The commutators can be found in the usual way, by decomposing tensor products
and imposing Jacobi identities.  They are the usual matrix commutators for
the $M^{\adt}{}_{\bdt}$ and $N^{\abar}{}_{\bbar}$ among themselves plus
$[M,N]=0$ as well as
\begin{eqnarray}
\nonumber
[M,Y]^{{\abar}{\adt}{\bdt}{\cdt}} =
3 \cdot Y^{\abar{\ddt}({\adt}{\bdt}}M^{\cdt)}{}_{{\ddt}}, \;
[Y,{Y'}]^{\adt}{}_{\bdt} &=& 
({Y'}^{{\abar}{\adt}{\cdt}{\ddt}}Y_{{\abar}{\bdt}{\cdt}{\ddt}}
  - Y^{{\abar}{\adt}{\cdt}{\ddt}}{Y'}_{{\abar}{\bdt}{\cdt}{\ddt}}),\\
\,[N,Y]^{{\abar}{\adt}{\bdt}{\cdt}} =
N^{\abar}{}_{\cbar}Y^{\cbar{\adt}{\bdt}{\cdt}}, \;
[Y,{Y'}]^{\abar}{}_{\bbar} &=& (
 {Y'}^{{\abar}{\adt}{\bdt}{\cdt}}Y_{\bbar{\adt}{\bdt}{\cdt}}
-Y^{{\abar}{\adt}{\bdt}{\cdt}}{Y'}_{\bbar{\adt}{\bdt}{\cdt}} ),
\label{HaupttextG2so3so3relations}
\end{eqnarray}
where group indices are lowered by contraction with the rightmost
index of totally antisymmetric $\varepsilon_{\adt\bdt}$ or
$\varepsilon_{\abar\bbar}$ with $\varepsilon_{12}=+1$. 
Examination of the Killing form $24\cdot\Tr(MM')+8\cdot\Tr(NN')
-16\cdot Y_{{\abar}{\adt}{\bdt}{\cdt}}{Y'}^{{\abar}{\adt}{\bdt}{\cdt}}$
shows that this defines the maximally non-compact $\mathfrak{g}_{2(+2)}$
if the generators are real and the coefficients satisfy symplectic
reality conditions 
$(M^{\adt}{}_{\bdt})^*=(M^*)_{\adt}{}^{\bdt}
=-M_{\adt\ddt}\varepsilon^{{\ddt}{\bdt}}, 
(N^{\abar}{}_{\bbar})^*= -N_{\abar\dbar}
\varepsilon^{\dbar{\bbar}}$
and 
$(Y^{{\abar}{\adt}{\bdt}{\cdt}})^*= -Y_{{\abar}{\adt}{\bdt}{\cdt}}$
(adopting the convention by which conjugation automatically shifts
index positions).

Next, for the realization of the different $\mathfrak{so}(3)$-representations.
Denoting the fundamental indices of $\mathfrak{so}(3)_F$ by $i,j,\ldots$,
the assignment of representations leads to an index structure 
$\phi^{\abar\adt\bdt\cdt}$ for the scalars, $\Psi^{i\abar}$ for gravitino and
supersymmetry parameter, and $\chi^{i\adt\bdt\cdt}$ for the matter fermions.
The action of infinitesimal $\mathfrak{so}(3)$-transformations is fixed by
linearity and by demanding for each two such transformations $X,Y$ that
$[\delta_X,\delta_Y]=\delta_{[Y,X]}$.  To ensure consistency, both the 
$\mathfrak{so}(3)_F$ coefficients and the fields inherit the symplectic
reality condition, 
$\nonumber (\varphi^*)_{\bar{a}\,\dot{a}\dot{b}\dot{c}}=-
\varphi_{\bar{a}\,\dot{a}\dot{b}\dot{c}}, (\chi^*)_{i\,\dot{a}\dot{b}\dot{c}}
= - \chi_{i\,\dot{a}\dot{b}\dot{c}}, (\Psi^*)_{i\bar{a}} = -\Psi_{i\bar{a}}.$
On the fermionic side, this makes fully contracted products of 
(anticommuting) spinors symmetric, with Clifford conjugation as their
adjoint, e.g.
$(\bar{\chi}_{i\,\dot{a}\dot{b}\dot{c}}\gamma^{\mu_1\cdots\mu_m}\zeta^{i\,\dot{a}\dot{b}\dot{c}})=(-)^{m(m+1)/2}(\bar{\zeta}_{i\,\dot{a}\dot{b}\dot{c}}\gamma^{\mu_1\cdots\mu_m}\chi^{i\,\dot{a}\dot{b}\dot{c}})$.
Introducing connection coefficients 
and $Q_{\mu}{}^{\abar}{}_{\bbar}$ and $Q_{\mu}{}^{\adt}{}_{\bdt}$ for the
$\mathfrak{so}(3)_B$ and $\mathfrak{so}(3)_2$, respectively, we can define
the action of a derivative covariant under these local symmetries,
$(D_{\mu}(Q)\Psi)^{i\bar{a}} =
\pa_{\mu}\Psi^{i\bar{a}}+Q_{\mu}{}^{\bar{a}}{}_{\bar{b}}\Psi^{i\bar{b}}$ 
and corresponding expressions for the other fields; replacing
$\pa_{\mu}$ by the Lorentz-covariant 
$D_{\mu}(\omega)=\pa_{\mu}+\frac14\gamma^{\alpha\beta}\omega_{\mu\alpha\beta}$,%
we obtain a derivative $D_{\mu}(\omega,Q)$ that is gauge- as well as Lorentz-covariant.

After these preparations, we can find the Lagrangian.  We restrict ourselves
to the terms that are necessary for the comparison with the dimensional
reduced theory, omitting quartic or higher fermionic terms in the Lagrangian.
Starting with the fields' standard kinetic terms and supersymmetry 
variations, supersymmetry demands the inclusion of a Noether term and
fixes ambiguities of the relative constants, with the resulting
Lagrangian
\begin{eqnarray}
\nonumber
\lag &=& e\bigg\{ -\frac{1}{4\kappa^2}\,\calR 
-\frac{i}{2}
(\overline{\Psi}_{\mu i\abar}\gamma^{\mu\nu\rho}
D_{\nu}(\omega,Q)\Psi_{\rho}^{i\abar})
-\frac{1}{2\kappa^2}g^{\mu\nu}(P_{\mu})_{\abar\adt\bdt\cdt}
       (P_{\nu})^{\abar\adt\bdt\cdt}\\[0.5em]
&&\phantom{e\bigg\{}
-\frac{i}{4}
(\overline{\chi}_{i\adt\bdt\cdt}\gamma^{\mu}{D}_{\mu}(\omega,Q)
  \chi^{i\adt\bdt\cdt})
+\frac{1}{\sqrt{2}}
 (\overline{\chi}_{i\adt\bdt\cdt}\gamma^{\rho}\gamma^{\mu}\Psi_{\rho}^{i\bbar})
 (P_{\mu})^{\abar\adt\bdt\cdt}\varepsilon_{\abar\bbar} \bigg\}
\hspace*{3em}
\label{3DFinalLagrangian}
\end{eqnarray}
invariant under the supersymmetry variations
\begin{eqnarray}
\nonumber
  \delta_S e_{\mu}{}^{\alpha} &=& i\kappa^2(\overline{\epsilon}_{i\bar{a}}
\gamma^{\alpha}\Psi_{\mu}^{i\bar{a}})\\
\nonumber
  \delta_S\Psi_{\mu}{}^{i\bar{a}} &=& 
   -(D_{\mu}(\omega,Q)\epsilon)^{i\bar{a}}
   -(\Sigma_S)^{\bar{a}}{}_{\bar{b}}
      \Psi^{i\bar{b}}\\
\nonumber\delta_S\chi^{i\dot{a}\dot{b}\dot{c}} &=&
\sqrt{2}i
\gamma^{\mu}(P_{\mu})^{\bar{a}\dot{a}\dot{b}\dot{c}}
\varepsilon_{\bar{a}\bar{b}} \epsilon^{i\bar{b}}
- 3\,
\chi^{i\dot{d}(\dot{a}\dot{b}}(\Sigma_S)^{\dot{c})}{}_{\dot{d}}\\
\nonumber
\delta_S(P_{\mu})^{\abar\adt\bdt\cdt} &=& 
-\frac{\kappa^2}{\sqrt{2}}\varepsilon^{\abar\bbar}D_{\mu}(Q)
     (\overline{\epsilon}_{i\bbar}\chi^{i\adt\bdt\cdt})
     -3\,(P_{\mu})^{\abar\ddt(\adt\bdt}(\Sigma_S)^{\cdt)}{}_{\ddt}
     -(\Sigma_S)^{\abar}{}_{\bbar} (P_{\mu})^{\bbar\adt\bdt\cdt}\\
\nonumber
\delta_S(Q_{\mu})^{\adt}{}_{\bdt} &=& D_{\mu}(Q)\Sigma_S{}^{\adt}{}_{\bdt}
-\frac{\kappa^2}{\sqrt{2}}
(P_{\mu})_{\abar\fdt\cdt\ddt}
       (2\delta^{\adt}_{\edt}\delta^{\fdt}_{\bdt}
            -\delta^{\adt}_{\bdt}\delta^{\fdt}_{\edt})
       \varepsilon^{\abar\bbar}(\overline{\epsilon}_{i\bbar}
                   \chi^{i\edt\cdt\ddt})\\
\delta_S(Q_{\mu})^{\abar}{}_{\bbar} &=& D_{\mu}(Q)\Sigma_S{}^{\abar}{}_{\bbar}
-\frac{\kappa^2}{\sqrt{2}}
(P_{\mu})_{\ebar\adt\bdt\cdt}
       (2\delta_{\bbar}^{\ebar}\varepsilon^{\abar\dbar}
            -\delta^{\abar}_{\bbar}\varepsilon^{\ebar\dbar})
       (\overline{\epsilon}_{i\dbar}
                   \chi^{i\adt\bdt\cdt}),
 \label{3DFinalSusy}
\end{eqnarray}
where $\Sigma_S$ is the highly non-linear expression 
$\Sigma_S = -\tanh (1/2\cdot\ad_{\varphi})\:\varphi$ that is a consequence
of describing the variation of group-valued objects in terms of 
algebra-valued objects \cite[sect.\ 2]{Marcus:1983hb}.  The coset-specific
properties come into play in that the variation of the Rarita-Schwinger 
term contains $[D_{\mu}(Q),D_{\nu}(Q)]$, to be cancelled by a term 
proportional to $[P_{\mu},P_{\nu}]$ arising from the variation of the
Noether term with respect to $\chi$.

\section{The dimensional reduction from five to three dimensions}

The stage is now set for the dimensional reduction
from which we mean to recover the model
found in the previous section.  Starting point is the minimal
$\mathcal{N}=2$ supergravity in five dimensions 
\cite{Cremmer:1980gs,Chamseddine:1980sp}.  We choose a ``mostly minus''
metric with the gravitini symplectic-Majorana spinors, denote curved
indices by $M,N,P,\ldots$ and flat indices by $A,B,C,\ldots$ and
define tensorial epsilon symbols, obtained from an all-flat
$\epsilon_{12345}=+1$ by applications of vielbein and metric.
The fields in question are vielbeins $E_M{}^A$, gravitini
$\Psi^i_M$ and a one-form field $A_M$; the Lagrangian is
\begin{eqnarray}
\nonumber
\lag_{5|2} &=& -\frac{1}{4\kappa^2}E\calR
-\frac14 EF^2
-\epsilon_3\frac{1}{6\sqrt{3}}(E\varepsilon^{MNPQR})
\kappa A_MF_{NP}F_{QR}\\
\nonumber
&&+\frac{i}{2}E(\overline{\Psi}_{Mi}\Gamma^{MNP}D_N(\omega)\Psi^i_P)
+\frac{\sqrt{3}}{8}i\kappa EF_{MN}
[2(\overline{\Psi}_i^M\Psi^{Ni})
+(\overline{\Psi}_{Pi}\Gamma^{MNPQ}\Psi_Q^i)]
\label{5dSugraLagrangian}
\end{eqnarray}
(with quartic fermionic terms omitted and with
$F_{MN}:=2\pa_{[M}A_{N]}$ the field strength), and it is invariant under
supersymmetry variations 
\begin{eqnarray}
\delta E_M{}^A = i\kappa^2
(\overline{\epsilon}_i \Gamma^A\Psi_M^i), \quad
\delta \Psi^i = \hat{D}(\omega,F)\epsilon^i  \quad \mbox{and} \quad
\delta A = \frac{\sqrt{3}}{2}\:i\kappa\:
(\overline{\epsilon}_i\Psi^i),
\label{d5N2FinalSusyVariation}
\end{eqnarray}
where the supercovariant derivative $\hat{D}$ is defined by
\begin{equation}
\hat{D}_M(\omega,F)\epsilon := D_M(\omega)\epsilon
 + \frac{1}{4\sqrt{3}}\kappa
(\Gamma_{ABC} + 4
\eta_{BC}\Gamma_{A})F^{AB}E_M{}^C\epsilon.
\label{d5N2SupercovariantDerivative}
\end{equation}
The initial steps of the dimensional reduction to three dimensions 
are fairly generic 
\cite{Cremmer:1979up,Cremmer:1980gs,Julia:1980gr,deWit:1984vq}:  
Exploiting part of the Lorentz gauge freedom, certain vielbein components
can be gauged to zero, allowing the decomposition
\begin{equation}
E_M{}^A = 
\left(\begin{array}{c@{\hspace*{6pt}}c}
\Delta^{-1}\; e'_{\mu}{}^{\alpha} & B_{\mu}{}^m e_m{}^a\\[6pt]
\mbox{\large 0} & e_m{}^a
\end{array}
\right),
\label{EMASplit}
\end{equation}
where $\Delta=\mathrm{det}\,e_m{}^a$ is a Weyl scaling factor.
Here and in the following, curved indices are decomposed in the manner
$M=(\mu,m)$, with $\mu$ a three-dimensional space-time index and $m$ 
an index in the two-dimensional internal space; flat indices are
split analogously as $A=(\alpha,a)$.  We adopt the convention of splitting
fields such as $F$ or the gravitino starting from their (all-lowered)
flat-indexed form; curved-index versions are then obtained by application 
of the component vielbeins $e'_{\mu}{}^{\alpha}$ and $e_m{}^a$.

The dimensional split of the anholonomy coefficients leads to
non-zero expressions
\begin{eqnarray}
\nonumber \Omega_{\alpha\beta}{}^{\gamma} &=& \Delta \left[
\Omega'_{\alpha\beta}{}^{\gamma} +2 
e'_{[\alpha}{}^{\mu}\delta_{\beta]}^{\gamma} \;(\pa_{\mu}\ln\Delta)\right]\\ 
\nonumber \Omega_{\alpha\beta}{}^c &=& \Delta^{2}\:\Omega'_{\alpha\beta}{}^c \\
\Omega_{a\beta}{}^{c} &=& \Delta\: \Omega'_{a\beta}{}^{c}
\label{SplitAnholonomyDurchAnholonomyPrime}
\end{eqnarray}
using $\Delta$-independent primed coefficients defined as 
\begin{eqnarray}
\nonumber \Omega'_{\alpha\beta}{}^{\gamma} &:=& 
-2 e'_{[\alpha}{}^{\mu}
e'_{\beta]}{}^{\nu}\:\pa_{\mu}(e'_{\nu}{}^{\gamma})\\
\nonumber \Omega'_{\alpha\beta}{}^c &:=& -
e'_{\alpha}{}^{\mu}e'_{\beta}{}^{\nu}\: G_{\mu\nu}^n e_n{}^c\\
  \Omega'_{a\beta}{}^c &:=& e'_{\beta}{}^{\nu}e_a{}^m,
(\pa_{\nu}e_m{}^c) 
\label{SplitAnholonomyPrime}
\end{eqnarray}
where $G_{\mu\nu}^n:=\pa_{\mu}B_{\nu}{}^n-\pa_{\nu}B_{\mu}{}^n$ is the
field strength of the Kaluza-Klein vector field.

On the fermionic side, the split is dimension-specific.  The five-dimensional
gamma matrices $\Gamma^A$ are split into $\Gamma^{\alpha}= 
\gamma^{\alpha}\hat{\Gamma}^{v}$ and $\Gamma^a=\hat{\Gamma}^a$,
with $\gamma^{\alpha}$ the three-dimensional matrices and $\hat{\Gamma}^a$
those of the two-dimensional internal space with signature $(-,-)$; 
for concreteness, we choose 
$\gamma^0=i\epsilon_3\sigma_2,
\gamma^1=\sigma_1, 
\gamma^2=\sigma_3, \GamHat^1=i\sigma_1, \GamHat^2=i\sigma_3$ and use
the abbreviation $\GamHat^v:=\GamHat^1\GamHat^2$.  The internal matrices
$\GamHat^a$ are $\mathfrak{so}(2)$-gamma matrices, but it is straightforward 
to use them to generate $\mathfrak{so}(3)$: The generators $\GamHat^r$,
$r=1,2,3$, where  $\GamHat^r=(\GamHat^a,\GamHat^v)$,
satisfy $\hat{\Gamma}^{r} \hat{\Gamma}^{s} = \varepsilon ^{rst}
\hat{\Gamma}_{t} -\delta^{rs},$ 
with $\varepsilon ^{123} = -1$, the $\mathfrak{su}(2)$ algebra.  
In addition, the reality condition
$((\GamHat^r)^{\abar}{}_{\bbar})^* = -\varepsilon_{\abar\cbar}\:
(\GamHat^r)^{\cbar}{}_{\dbar}\:\varepsilon^{\dbar\bbar}$ (with $SO(2)$-spinor
indices $\abar,\bbar,\ldots$) which these matrices
inherit is the same that is needed for the $\mathfrak{so}(3)$-generators in
the $\mathfrak{g_{2(+2)}}$ decomposition introduced above.  From Fierz
identities for the $\GamHat$'s Clifford algebra, one can derive useful 
relations such as 
\begin{eqnarray}
\nonumber
(\hat{\Gamma}^r)^{\abar}{}_{\bbar}(\hat{\Gamma}_r)^{\cbar}{}_{\dbar}
&=&2\delta^{\abar}_{\dbar}\delta^{\cbar}_{\bbar}
-\delta^{\abar}_{\bbar}\delta^{\cbar}_{\dbar}, \\
\nonumber
\varepsilon_{rst}
(\hat{\Gamma}^s)^{\abar}{}_{\bbar}(\hat{\Gamma}^{t})^{\cbar}{}_{\dbar} 
&=& \delta
^{\abar}_{\dbar} (\hat{\Gamma}_{r})^{\cbar}{}_{\bbar} - \delta
^{\cbar}_{\bbar} (\hat{\Gamma}_{r})^{\abar}{}_{\dbar} \mbox{ and } \\
(\GamHat^a)^{\ebar}{}_{[\bbar}(\GamHat^v)^{\dbar}{}_{{\gbar}]} 
&=&(\GamHat^v\GamHat^a)^{\ebar}{}_{[\bbar}\delta^{\dbar}_{\gbar]},
\label{so3moregeneralfromfierz}
\end{eqnarray}
to be exploited later on.  Let us note one consequence of these relations,
namely that, using a flat $Spin(2)$-invariant metric $\delta_{\abar\bbar}$
to lower indices, one can derive
\begin{equation}
(\GamHat^a)_{\abar\bbar}(\GamHat_a)_{\cbar\dbar}
+(\GamHat^a)_{\dbar\hspace*{1pt}\bbar}(\GamHat_a)_{\cbar\abar}
=2\delta_{\abar\dbar}\delta_{\bbar\cbar}.
\end{equation}
This is a Clifford relation, but with contraction over vector instead of
spinorial indices, corresponding to identical relations for $SO(8)$ which
are associated with the triality property of that group
and are used in the construction of the $E_{8(+8)}/SO(16)$ model.

As for the five-dimensional gravitino, and adopting the convention to
suppress (five- and three-dimensional) space-time spinor indices, the 
first decomposition is of $\Psi^i_A$ into a three-dimensional 
spin-3/2 fermion $\Psi^{i\abar}_{\alpha}$ and a spin-1/2 fermion
$\Psi^{i\abar}_{a}$.  $\abar$ can be promoted to an 
$\mathfrak{su}(2)\iso\mathfrak{so}(3)$-index: As far
as internal spinor indices are concerned, the spinor product's adjoint is
Hermitian conjugation, with an invariance group $U(2)$.  The symplectic
reality condition restricts this to $SU(2)$.
However, there are problems:  This decomposition
leads both to a non-standard form for the threebein's supersymmetry
variations and to mixed first-derivative terms between spin-3/2 and
spin-1/2 degrees of freedom.  The remedy is a redefinition
$\Psi'{}_{\mu}^{i\abar} =
\Delta^{-1/2}[\gamma_{\beta}(\GamHat^{v}\GamHat^a)^{\abar}{}_{\bbar}\Psi_a^{i\bbar}-\Psi^{i\abar}_{\beta}]e'_{\mu}{}^{\beta}$;
with this field as the three-dimensional gravitino and $e'_{\mu}{}^{\beta}$
as vielbein, the dimensional reduction reproduces both the gravitino kinetic 
term (with the mixed terms now absent) and the vielbein 
supersymmetry variation given in (\ref{3DFinalLagrangian}) and
(\ref{3DFinalSusy}), respectively.  From this, it is tempting to identify
the gravitino's index $\abar$ as an index of $\mathfrak{so}(3)_B$.
However, matters are more complicated, which can be seen by considering
the spin-1/2 fields.  To start with, they have
the index structure $\Psi_a^{i\abar}$, where $\abar$ can be promoted to an
$\mathfrak{so}(3)$ index, as before.  However, from the construction in section
\ref{3DSection} we know that the matter fermions transform non-trivially
only with respect to $\mathfrak{so}(3)_2$ and $\mathfrak{so}(3)_F$.
This apparent problem can be resolved once it is realized that in similar
situations, notably in the $E_8$ case, dimensional reduction leads to 
models in which the enhanced local symmetry is gauge-fixed, with
expressions e.g.\ for the $P_{\mu}$ and $Q_{\mu}$ that are not explicitly 
covariant under that symmetry.  Our case is analogous in that, apparently, the
dimensionally reduced model ``sees'' only the diagonal
$\mathfrak{so}(3)$ subgroup of $\mathfrak{so}(3)_B\times\mathfrak{so}(3)_F$.
To restore the enhanced symmetry, that diagonal group needs to be
disentangled into its $\mathfrak{so}(3)_B\times\mathfrak{so}(3)_F$ 
parts, using the model developed in section \ref{3DSection} as a guide.
This promotes the $\abar$ index of $\Psi'{}_{\mu}^{i\abar}$ to an 
$\mathfrak{so}(3)_B$ index, and the internal spinor index of the
spin-1/2 fields to an $\mathfrak{so}(3)_2$ index.
For the latter, the kinetic term should have
the same simple form as shown in (\ref{3DFinalLagrangian}).  This can 
be achieved by exploiting the freedom to re-define $\Psi^{i\adt}_a\to
\Psi^{i\adt}_a+(\GamHat_a\GamHat^c)^{\adt}{}_{\cdt}\Psi^{i\cdt}_c$.
All in all, new matter fermion fields defined as
\begin{equation}
\chi^{i\adt\bdt\cdt} = \Delta^{-1/2}\Psi^{i\edt}_c[
\delta^c_d\delta^{(\adt}_{\edt}
+(\GamHat_d\GamHat^c)^{(\adt}{}_{\edt}](\GamHat^d)^{\bdt}{}_{\ddt}
\varepsilon^{\cdt)\ddt}
\end{equation}
have the required properties.  It can be checked directly that
the fermions thus defined
have inherited the correct symplectic reality condition.
With these preparations, the terms quadratic
in $\Psi'$ that are obtained from the five-dimensional Rarita-Schwinger term
combine into the three-dimensional enhanced Rarita-Schwinger term with
gauge-fixed connection coefficients
\begin{eqnarray}
\nonumber
Q_{\nu}{}^{\bar{a}}{}_{\bar{b}} &:=&
-\frac{1}{4}e'_{\nu}{}^{\alpha}\bigg\{
 (\GamHat^{de})^{\bar{a}}{}_{\bar{b}}\Omega'_{\alpha de}
+\epsilon_3\Delta(\GamHat^v\GamHat^e)^{\bar{a}}{}_{\bar{b}}
\Omega'_{\alpha e}\\
&&
\phantom{ -\frac14e'_{\nu}{}^{\alpha}\bigg\{}
+2\sqrt{3}\kappa\Delta^{-1}\left[
\epsilon_3(\GamHat^v)^{\bar{a}}{}_{\bar{b}}F_{\alpha}
+(\GamHat^d)^{\bar{a}}{}_{\bar{b}}F_{\alpha d}\right]
\bigg\},
\label{d5NewabarbbarConnectionCoefficients}
\end{eqnarray}
while the corresponding connection coefficients in the kinetic term of
the spin-1/2 fermions turn out to be
\begin{eqnarray}
\nonumber
Q_{\nu}{}^{\adt}{}_{\ddt} &=&-\frac{1}{4}e'_{\nu}{}^{\beta}
\bigg[
\phantom{+}(\GamHat^v)^{\adt}{}_{\ddt}(\varepsilon^{cd}\Omega'_{\beta cd}
+{\scriptstyle \frac{2}{\sqrt{3}}}
     \epsilon_3\kappa\Delta^{-1} F_{\beta})\\
&&
\phantom{\frac{1}{8a_r}(1+i)e'_{\nu}{}^{\beta}
\bigg[}
-(\GamHat^c)^{\adt}{}_{\gdt}(\Delta\epsilon_3(\GamHat^v)^{\gdt}{}_{\ddt}
                            \Omega'_{\beta c}
-{\scriptstyle \frac{2}{\sqrt{3}}}\kappa\Delta^{-1}\delta^{\gdt}_{\ddt}
F_{\beta c}) \bigg].                         
\label{d5adtbdtConnectionCoefficients}
\end{eqnarray}
The mixed term containing both $\Psi'$ and $\chi$, compared with the
Noether term in the three-dimensional target lagrangian, yield 
\begin{eqnarray}
\nonumber 
(P_{\mu})^{\bar{a}\ddt\edt\fdt} &=&
-\frac{i}{2\sqrt{2}}\: \varepsilon^{\bar{a}\bdt}
e'_{\mu}{}^{\alpha}\bigg\{
\frac12\Delta
\epsilon_3\Omega'_{\alpha c}
 \left[\delta_{\bdt}^{(\ddt}(\GamHat^c)^{\edt}{}_{\gdt}\varepsilon^{\fdt)\gdt}
+(\GamHat^v\GamHat^c)^{(\ddt}{}_{\bdt}(\GamHat^v)^{\edt}{}_{\gdt}
\varepsilon^{\fdt)\gdt}\right]\\
\nonumber &&
\phantom{\varepsilon^{\bar{a}\bdt}
e'_{\mu}{}^{\alpha}\bigg\{}
+%
\left[
(\GamHat^v)^{(\ddt}{}_{\gdt}\delta_{\bdt}^{\edt}\varepsilon^{\fdt)\gdt}
\Omega'_{\alpha d}{}^d
+(\GamHat^v\GamHat^c)^{(\ddt}{}_{\bdt}
(\GamHat^b)^{\edt}{}_{\gdt}\varepsilon^{\fdt)\gdt}\Omega'_{\alpha(bc)}
\right]\\[0.5em]
\nonumber && \phantom{\varepsilon^{\bar{a}\bdt}
e'_{\mu}{}^{\alpha}}
+\sqrt{3}\kappa\Delta^{-1}\big[ F_{\alpha c}
(\GamHat^c)^{(\ddt}{}_{\gdt}
(\GamHat^v)^{\edt}{}_{\bdt}\varepsilon^{\fdt)\gdt}
+\epsilon_3F_{\alpha}
(\GamHat^v)^{(\ddt}{}_{\bdt}
(\GamHat^v)^{\edt}{}_{\gdt}\varepsilon^{\fdt)\gdt}\bigg\}.
\label{DimRed53PmuTerm}
\end{eqnarray}
By their index structure, it can be checked directly that these objects
transform under the proper representations of 
$\mathfrak{so}(3)_2\times\mathfrak{so}(3)_B$; by using the reality
conditions for the $\GamHat$, that they satisfy the proper reality conditions.

To complete the match, we present three consistency checks.  From 
(\ref{3DFinalSusy}), it follows that $Q_{\nu}{}^{\abar}{}_{\bbar}$ must
occur in the new gravitino's supersymmetry variations, while 
one should also be able to read off $(P_{\mu})^{\bar{a}\ddt\edt\fdt}$ 
from the matter fermion's supersymmetry variation.  Both expressions agree
with those derived above.  The final check is the match of the dimensionally
reduced bosonic Lagrangian with the sigma model kinetic term in terms
of the $P_{\mu})^{\bar{a}\ddt\edt\fdt}$ of eq.\ (\ref{DimRed53PmuTerm});
the same check used in \cite{Mizoguchi:1998wv} for the 
$\frak{g}_2$-construction in terms of $\mathfrak{sl}(3)$ representations.
The sigma-model term is
\begin{eqnarray}
\nonumber -\frac{1}{2\kappa^2}g^{\mu\nu}(P_{\mu})_{\abar\adt\bdt\cdt}
(P_{\nu})^{\abar\adt\bdt\cdt}
&=& e'\bigg\{
-\frac{1}{16\kappa^2}(\pa_{\nu}\bar{g}_{mn})(\pa^{\nu}\bar{g}^{mn})
+\frac{1}{4\kappa^2}(\pa_{\nu}\ln\Delta)(\pa^{\nu}\ln\Delta)\\
&&
+\frac{\Delta^2 G^2}{16\kappa^2}
-\frac12(\pa_{\mu}A_m)(\pa^{\mu}A_n)\bar{g}^{mn}
-\frac14\Delta^{-2}(F')^2\bigg\},
\label{ePPKineticBosonicTerms}
\end{eqnarray}
while the reduction of the bosonic terms gives the three-dimensional 
Einstein-Hilbert term plus
\begin{eqnarray}
\nonumber
e'\bigg\{\frac{\Delta^2G^2}{16\kappa^2}
+\frac{1}{16\kappa^2}(\pa^{\mu}\bar{g}_{mn})(\pa_{\mu}\bar{g}^{mn})
-\frac{1}{4\kappa^2}(\pa^{\mu}\ln\Delta)(\pa_{\mu}\ln\Delta)
-\frac14\Delta^{-2}(F')^2\\
+\frac12\bar{g}^{mn}(\pa_{\mu}A_m)(\pa_{\nu}A_n)
-\epsilon_3\kappa\frac{1}{3\sqrt{3}}
\:\varepsilon^{\mu\nu\rho}\varepsilon^{mn}A_m(\pa_{\rho}A_n)
[3F'_{\mu\nu}+\Delta^2G_{\mu\nu}^pA_p]\bigg\}. \qquad
\label{5dDimRedBosonicLagrangian}
\end{eqnarray}
That there is no match comes as no surprise, as it is a general feature of
hidden symmetries to become manifest only upon dualization of appropriate
dimensionally-reduced $p$-forms.  In this case, the objects that allow
dualization are the Kaluza-Klein field strength $G_{\mu\nu}^m$, dual to 
two scalars $\xi_m$, and a composite ``field strength''
$\tilde{F}_{\mu\nu} := \Delta^{-2}F'_{\mu\nu} -G_{\mu\nu}^mA_m$, tailor-made
to fulfill the Bianchi identity and dual to a scalar $\varphi$.  After
dualization, the original Lagrangian (\ref{5dDimRedBosonicLagrangian})
plus the constraint terms becomes
\begin{eqnarray}
\nonumber
\hspace*{-1.5em}&&  -\frac12\Delta^{-2}g'{}^{\rho\lambda}
\left((\pa_{\rho}\varphi)-\frac{2}{\sqrt{3}}\epsilon_3
\Delta^2\kappa\varepsilon^{nr}A_n(\pa_{\rho}A_r)\right)\!\!\cdot\!\!
\left((\pa_{\lambda}\varphi)-\frac{2}{\sqrt{3}}\epsilon_3
\Delta^2\kappa\varepsilon^{nr}A_n(\pa_{\lambda}A_r)\right)\\
\nonumber
\hspace*{-1.5em}&&+2\kappa^2\Delta^{-2}
g'{}^{\rho\lambda}\phantom{\cdot\!\!}\left((\pa_{\rho}\varphi)A^m
-(\pa_{\rho}\xi_p)\bar{g}^{mp}
+\frac{2}{3\sqrt{3}}\epsilon_3\kappa A^m
\varepsilon^{pr}A_p(\pa_{\rho}A_r)\right)\hspace*{1em}\\
\hspace*{-1.5em}&& \phantom{+2\kappa^2\Delta^{-2}
g'{}^{\rho\lambda}}
\hspace*{-0.25em}\cdot\!\!\left((\pa_{\rho}\varphi)A_m
-(\pa_{\rho}\xi_m)\phantom{\bar{g}^{mp}}
+\frac{2}{3\sqrt{3}}\epsilon_3\kappa A_m
\varepsilon^{pr}A_p(\pa_{\rho}A_r)\right),
\end{eqnarray}
which is the same as the $P^2$-Lagrangian (\ref{ePPKineticBosonicTerms})
upon substitution of the dualized entities.  This completes our
cross-checks.

There are a number of possible directions for extending the present results.
The possibility of exploring ``exceptional geometries'' has already 
been mentioned; another interesting question would be to what
happens to the hidden symmetry if the $R$-symmetry or some subgroup
thereof is gauged \cite{DAuria:1981kq,Gunaydin:1985ak} (making contact
with recent, more general, studies of the possible gaugings
in three dimensions  \cite{Nicolai:2001sv}) or to study the case of
compactification on $AdS_3\times S^2$, for which the spectrum has
already been worked out in \cite{Fujii:1998tc}.

\section*{Acknowledgements}

We would like to thank Hermann Nicolai and Henning Samtleben for
stimulating discussions.  The work of M.\ P.\ was in part supported by 
Studienstiftung des deutschen Volkes, that of S.\ S.\ by the 
European Union under contract HPRN-CT-2000-00122.


\end{document}